
\input harvmac
\noblackbox

\def\pl#1#2#3{Phys. Lett. {#1}B (#2) #3}
\def\prl#1#2#3{Phys. Rev. Lett. {#1} (#2) #3}
\def\physrev#1#2#3{Phys. Rev. {D#1} (#2) #3}

\def\lfm{\smallskip\noindent\item}
\def\abs#1{\left| #1\right|}
 \Title{hep-ph/9511350,   UW/PT 95-18}
{\vbox{
\centerline{A Viable  Model  of Dynamical }
\smallskip
\centerline{Supersymmetry Breaking in the Hidden Sector}
 }}\bigskip
\centerline{ Ann E. Nelson }
 \smallskip
\centerline{ \it Department of Physics, Box 351560}
\centerline{\it University of Washington, Seattle, WA 98195-1560}
\bigskip
\baselineskip 18pt
\centerline{\bf Abstract}
\noindent
We propose a simple and natural model of dynamical supersymmetry breaking,
which could be
used as a mechanism for spontaneous supersymmetry breaking in a gravitationally
coupled hidden sector.  The gaugino masses in the visible sector are
naturally  of the same size as the squark, slepton, and gravitino masses.
\Date{11/95}
\newsec{Introduction}

Dynamical Supersymmetry Breaking (DSB) \ref\witten{
  E. Witten,   Nucl. Phys. B188 (1981) 513
 } is an attractive
explanation of the large hierarchy of scales introduced to describe
fundamental physics.  The usual minimal supersymmetric extension of the
 standard
model,   known as the MSSM, is based on the assumption   that supersymmetry is
spontaneously broken at a scale of
$\sim 10^{11}$ GeV in a ``hidden'' sector, with soft supersymmetry breaking
mass
terms of order
$\sim 10^2$ GeV arising in the visible sector from supergravitational effects.
The
MSSM scenario still requires   the input of a large  ($10^8$) hierarchy between
the
Planck and supersymmetry breaking scales, which could be explained by DSB. The
aim of this paper is to give an explicit example of   supersymmetry breaking in
the
hidden sector with the following attributes:
\lfm{1.} The supersymmetry breaking in the hidden sector is dynamical. The only
explicit scale in the theory is the Planck mass $m_P$, with any other scales
deriving from dimensional transmutation\foot{Except for a constant in the
superpotential needed to tune the cosmological constant to zero. We hope that
such tuning might eventually be described by effects within a nonsupersymmetric
effective theory of quantum gravity, and will not depend on the details of the
effective Lagrangian.}.
\lfm{2.} The supersymmetry breaking is described by an effective
supersymmetric gauge theory with a stable, nonsupersymmetric ground state at
field
strengths which are much smaller than
$m_P$.
\lfm{3.} Visible sector gaugino masses of the same size as the squark and
slepton masses can arise naturally, which translates into a requirement that
the hidden sector should contain a gauge singlet field with an F term as large
as any other F terms in the theory.

The reader may be sceptical about the necessity of this exercise as there are
already several proposals in the literature for hidden sector dynamical
supersymmetry breaking. However the existing models all fail to satisfy  either
the second or the third criterion\foot{While this work was in the process
of completion I learned about some overlapping independent work on the subject
of
hidden sector DSB by  R. Plesser and Y. Nir \ref\pn{R. Plesser, Y. Nir, private
communication }.}. Our rational for imposing these criteria is as
follows.

Most
proposed hidden sector DSB models have a ground state   at field strengths
which are
as large or larger than the Planck scale. Furthermore it is frequently stated
that
in these models supersymmetry is spontaneously broken by supergravitational
effects.
One should be able to redefine the field  content of these models to study
  the
effective theory of light excitations about the ground state, which in the
reformulated theory occurs for fields with small expectation values and
conventional
kinetic energy terms. This effective theory should  be supersymmetric and
should describe  spontaneous supersymmetry breaking at a scale well below the
Planck scale. Then there is a simple argument, due to Weinberg
\ref\weinberg{S. Weinberg,
\prl{48}{1982}{1776}} that in such an effective theory the supersymmetry
breaking will persist in the
flat space limit where the effects of supergravity are turned off.
In the presence of supergravity the scalar potential is
\ref\sugra{
  E. Cremmer, S. Ferrara, L. Girardello, A. Van Proeyen, Phys. Lett. 116B
(1982) 231; R. Barbieri, S. Ferrara, C.A. Savoy,   Phys. Lett. 119B (1982) 343
}
\eqn\sugrapot{\eqalign{V=&e^{8\pi K/m_P^2}
\left( (K_i^j)^{-1}
\left( W_i+8\pi K_i W/m_P^2 \right)
\left( W_j+8\pi K_j W/m_P^2 \right)^*
-24\pi|W|^2/m_P^2 \right) \cr
&+{1\over2}\sum_A \abs{K_i\,{t_A}^i_j\phi^j}^2\ .}} Here $W$ is the
superpotential, $K$
is the K\"ahler potential, and the $\phi^i$ are chiral superfields. If the
equations
\eqn\sugrafterms{\eqalign{&(W_i+8\pi K_i W/m_P^2)=0, \ {\rm all}\ i \cr
&   K_i\, {t_A}^i_j\phi^j=0\cr}} have a solution
then this solution is both supersymmetric and an extremum of the potential
\sugrapot, so supersymmetry is not broken. Weinberg argues that if a solution
for a
{\it globally} supersymmetric ground state exists
\eqn\superfterms{\eqalign{&W_i= 0\cr
&  K_i\,{t_A}^i_j\phi^j=0}} then a
small perturbation of this solution gives a solution to \sugrafterms\ which is
locally
supersymmetric, and which is a local minimum of the potential
\sugrapot.
Thus in order to obtain a model with local supersymmetry breaking one
should first find a globally supersymmetric theory without a
supersymmetric
ground
state. Therefore in this letter we focus on breaking global
supersymmetry spontaneously.
Supergravity
can then be treated as a  perturbation  which
  leads  to  small but important effects, such as soft supersymmetry
breaking
in
the visible sector.

It is well known \ref\susybreaking{P. Fayet, J. Iliopoulos,
Phys. Lett. B51 (1974) 461;  L. O'Raifeartaigh, Nucl. Phys. B96 (1975) 331;
} that global
supersymmetry breaking can occur at tree-level in models which introduce the
scale of supersymmetry breaking as a   parameter in the tree level
Lagrangian---and perhaps this is the correct low energy effective field
theory description of some nonperturbative effect arising from Planck scale
physics.

We find it more plausible that an effective theory of  DSB
explains the hierarchy between
the supersymmetry breaking and Planck scales. Viable examples of DSB have been
constructed where visible renormalizable gauge interactions communicate
supersymmetry
breaking radiatively to squarks, sleptons and gauginos \ref\dnns{M. Dine, A. E.
Nelson,
Y. Nir, Y. Shirman, to be published in Phys. Rev. D, hep-ph/9507378
}. However Affleck, Dine and Seiberg pointed out
\ref\ads{I. Affleck, M. Dine, N. Seiberg, Nucl. Phys. B256 (1985) 557
} that  in hidden sector DSB models it appears to be
difficult to give the ordinary gauginos a reasonable mass.  A gaugino mass term
must
come from the following term in the effective supersymmetric Lagrangian
\eqn\gauginomass{\int d^2\theta f(X_i/m_P)W_\alpha W^\alpha +{\rm h.c.}}
where $f$, the gauge kinetic function, is a holomorphic function of  $X_i$, the
chiral superfields. Supergravitational (and other nonrenormalizable effects
aring from
Planck scale physics) gives   squark and slepton masses of order the gravitino
mass
$m_{3/2}\sim (1/m_P)$. Phenomenologically viable gaugino masses should also
arise at
$\CO(1/m_P)$  and so
$f$ should contain a term linear in a chiral superfield which has an F term of
order the
supersymmetry breaking scale squared. Such a gauge kinetic function
can only be gauge invariant if this
superfield is a gauge singlet.  However   in known DSB models there are no
gauge
singlets participating in the DSB dynamics, and finding new DSB models,
especially ones
with gauge singlets, is nontrivial~\ref\constraints{ E. Witten,  Nucl .Phys.
B202
(1982) 253}.

One possibility \ref\bkn{T. Banks,
D. B. Kaplan, A. E. Nelson,  Phys. Rev. D49 (1994) 779, hep-ph/9308292
} is to   gauge a global symmetry of a known DSB
model\foot{Note that simply gauging a global symmetry of a DSB model never
restores supersymmetry, as argued in  ref. \ref\nelsei{ A. E. Nelson, N.
Seiberg, Nucl.
Phys. B416 (1994) 46,  hep-ph/9309299
}. }  and then use this new gauge group to feed
supersymmetry breaking radiatively to a gauge singlet. Then this gauge singlet
obtains an F-term at one loop, and one expects gaugino masses in the visible
sector to be
suppressed relative to squark and slepton masses by this loop
factor. It would be interesting
to know, however, whether such light gaugino
masses\foot{ There remains a phenomenologically viable window for such
light gaugino masses \ref\lightino{ G. R. Farrar,
hep-ph/9508292, hep-ph/9508291; J.L. Feng, N. Polonsky, S. Thomas,
hep-ph/9511324}. Other ways of extending the
visible sector to give the gauginos a mass were studied in
\ref\rsymmetry{ L.J. Hall, L. Randall, Nucl. Phys. B352 (1991) 289; M. Dine, D.
MacIntire, Phys. Rev. D46 (1992) 2594, hep-ph/9205227  }.}
are a necessary consequence
of hidden sector DSB.

In the next section we will show how in a modified version of a known DSB
model,  gauge
singlets do get F terms as large as
$\CO(M_s^2)\sim  m_{3/2}m_P$, and Planck scale physics can easily produce
ordinary
gaugino masses as large as the squark and slepton masses.

\newsec{A Hidden Sector Model of Dynamical Supersymmetry Breaking}
Consider the
gauge group SU(7) with matter content
\eqn\matter{A=21,\quad \bar F_i=\bar7, i=1,2,3\ ,}
($A$ is the two-index antisymmetric tensor).

The model was originally studied
by Affleck, Dine and Seiberg  (ADS) \ads. They
noted that no superpotential could be dynamically generated, and that in the
absence of a tree level superpotential   there is a classically
flat direction with
\eqn\flat{\eqalign{
&\bar F_1=(1/\sqrt3)\pmatrix{\phi& 0 &0& 0&0&0&0},\cr&
\bar F_2=(1/\sqrt3)\pmatrix{0&\phi&0&0&0&0&0},\cr
&\bar F_3=0,\cr &
{A^{\dag}} A={\rm diag}(|\phi^2|/6,|\phi^2|/6,0,0,0,0,0) \ ,}}
(global SU(3) rotations of this direction are also flat). For $\phi$
large,
the low energy effective theory includes SU(5) with a $\bar 5+ 10$, and a light
gauge singlet ``dilaton'' field $\phi$. The scale at which the SU(5) theory
becomes strong is
\eqn\lambdalight{\Lambda_5=\phi^{-4/13}\Lambda_7^{17/13}\ .} Since there is
strong
evidence that the SU(5) effective theory dynamically breaks supersymmetry at a
scale
$M_s\sim\Lambda_5$ \ref\noncalc{I. Affleck, M. Dine, N. Seiberg, Phys. Lett.
137B
(1984) 187; H. Murayama, Phys. Lett. B355 (1995) 187,  hep-th/9505082; E.
Poppitz, S.
P. Trivedi, preprint EFI-95-44 (Jul 1995) hep-th/9507169
 } we expect that  the dilaton slides
off to infinity and the theory has no ground state. ADS then noted that if one
adds
a superpotential, e.g.
\eqn\treesuper{W=A \bar F_2 \bar F_3\ ,} then there are no classically flat
directions,
and there is a stable  vacuum with supersymmetry spontaneously broken.
Complete quantitative  understanding
of the dynamics of the supersymmetry breaking and the particle spectrum
involves strong
interactions and is not ``calculable''.

Let us now take this SU(7) theory, (without the superpotential \treesuper),
and add 3 gauge singlet fields
\eqn\xfields{X_i,i=1,2,3} with (nonrenormalizable) superpotential couplings
\eqn\nonrensuper{W=(1/m_P)\sum_{i, j, k=1,2,3}\epsilon_{ijk}\lambda_i X_i A
\bar F_j
\bar F_k\ .}  This theory has no classical flat directions
involving SU(7)-colored fields although the $X$'s are undetermined classically
when SU(7)
is unbroken. Note that that mass and self coupling terms for the $X$'s
can be eliminated by an R
symmetry under which the
$X$'s have zero charge, and this R symmetry allows the gauge kinetic functions
to depend on the $X_i$'s\foot{We thank Michael Dine for this
comment.}. A dimension four term such as \treesuper\ can  be
made  zero by shifting the $X_i$ fields.

Now there is competition between two couplings. The SU(7) dynamics would like
$\phi$
to be large in order to reduce the scale of
supersymmetry breaking. However due to the superpotential \nonrensuper,
assuming
the $\lambda_i$'s are all of order one,  there is always an F term for
one linear combination of the $X$'s
which has size
$\phi^3/m_P $.  The potential is minimized for $\phi$ of order
\eqn\vevscale{\phi\sim\Lambda_7^{(34/47)}m_P^{(13/47)}\ .}
which gives a supersymmetry breaking at a scale \eqn\dsbscale{
M_s\sim \phi^3/m_P\sim
\Lambda_7^{(51/47)}m_P^{(-4/47)}\ .}  In Appendix A we show that for
$\phi$ non zero,   the potential also
eliminates the flat directions for the scalar components of
$X_i$. The tree level potential    gives the $X$ scalars a mass of order
\eqn\xmass{M_X\sim M_s^2/\phi\ .}

The F term for one linear combination of the gauge singlets $X_i$ is of the
same order
as the supersymmetry breaking scale   and provided that the ordinary gauge
kinetic function contains a term like \gauginomass\ which is linear in the
$X$'s  the
ordinary gauginos will get a mass of the same size as the supergravity-induced
squark and
slepton masses. If $\Lambda_7$ is about $\sim 10^{11}$ GeV,   then
the  gravitino mass will be of order
$10^2$ GeV. Superpartners in the visible sector will have masses of order the
weak scale. The   hidden sector scalars $X$ have masses around $3\times 10^7$
GeV and
do not lead to any ``Polonyi-type'' \nref\mod{ B. de Carlos, J.A. Casas, F.
Quevedo, E.
Roulet,
Phys. Lett. B318 (1993) 447, hep-ph/9308325;
 T. Banks, M. Berkooz, P.J. Steinhardt, Phys. Rev. D52 (1995) 705,
hep-th/9501053; T.
Banks, M. Berkooz, S.H. Shenker, G. Moore, P.J.
Steinhardt, hep-th/9503114
}\refs{\bkn,\mod }
cosmological problems\foot{The analysis of hidden sector scalar masses in
ref.~\bkn\ does
not apply here  or to any model where the flat directions   are lifted by
nonrenormalizable terms in the superpotential.}. There is  a Goldstone boson,
due to
spontaneous R symmetry breaking at a scale
$\phi\sim 10^{13}$ GeV. This Goldstone boson could play  the role of an
invisible axion
\ref\axion{R. Peccei, H. Quinn, \prl{38}{1977}{1440}; \physrev{16}{1977}{1791};
S. Weinberg, \prl{40}{1978}{223};
F. Wilczek, \prl{46}{1978}{279}; J. E. Kim,  Phys. Rev. Lett. 43 (1979) 103;
M.A. Shifman, A.I. Vainshtein, V.I. Zakharov, Nucl. Phys. B166 (1980) 493;
M. Dine, W. Fischler, M. Srednicki, \pl{104B}{1981}{199}
} in the
ordinary QCD sector and   solve the strong CP problem.
However we expect to softly break the R symmetry by
explicitly adding a constant to the superpotential in order to
fine-tune the cosmological constant to zero---this explicit symmetry breaking
will give
the axion a large mass
\ref\baggeretal{ J. Bagger, E. Poppitz, L. Randall, Nucl. Phys. B426 (1994) 3,
hep-ph/9405345  }, rendering the axion irrelevant both for
cosmology and for the strong CP problem.

Now that we have an explicit example,  we can present an algorithm for
finding models of DSB with gauge singlets. Take a chiral gauge theory which
is known to have D flat directions and no supersymmetric ground state, add  a
number of
gauge singlet superfields equal to the number of flat directions, and add  a
tree level
superpotential which couples the gauge singlets to the gauge invariant
polynomials of
chiral superfields which parametrize the flat directions \ref\poly{D. Mumford
and J.
Fogarty, {\it Geometric Invariant Theory} (Springer, 1982); M.A. Luty, W.
Taylor IV,
preprint  MIT-CTP-2440 (June, 1995),  hep-th/9506098}, so that the only
remaining
classically flat directions have unbroken gauge symmetry. Most of the resulting
models
do not have a stable supersymmetry breaking ground state, but  our example is
not
the unique DSB model which could serve as a supersymmetry breaking hidden
sector.

\newsec{A    Dilaton Problem?}
We have not included
the   dilaton   predicted by string theory, and so do not have the ``dilaton
problem'' of
a  supersymmetric ground state
\ref\dilaton{ M. Dine, N. Seiberg, Phys. Lett. 162B (1985) 299  } at infinite
dilaton
field strength. If we assume our model arises as the low energy limit of a
string theory
we must also assume that some other mechanism    (not associated with
supersymmetry breaking) stabilizes the superstring dilaton\foot{
 For instance  in some  models with two or more gauge groups and no
matter, there is a supersymmetric minimum with a stable dilaton,
\ref\racetrack{N.V.
Krasnikov,    Phys. Lett. B193 (1987) 37,    T.R. Taylor  Phys.
Lett. B252 (1990) 59}. We could simply add a such a dilaton stabilizing sector
to our
theory.}. One might worry\foot{We thank M. Dine for correspondance on this
point.} that
since our
$X_i$ superfields  have dilaton-like couplings to gauge fields, as in
eq.~\gauginomass,
there is an   direction associated with some $X_i\rightarrow
\infty$ in which supersymmetry tends to be restored. However,   unlike in the
case of
the superstring dilaton, we have no theoretical reason to severely constrain
the $X_i$   couplings,  and so there
is no way to decide, (without input from short distance physics) whether the
gauge
couplings get stronger or weaker as $X_i\rightarrow\infty$, and whether
supersymmetry is
 restored in this limit. In any case,  the contribution to
the mass squared of the $X_i$ particles  near the origin of field space from
the
dependence of the gauge coupling constant on the vev $\vev{X_i}$ is of order
\eqn\xmassgaugecont{
\Delta M_{X_i}^2\sim {\partial^2 M_s^4\over (\partial X_i)^2} \Big|_{X_i=0}\sim
{\partial^2
\Lambda_5^4\over (\partial X_i)^2}\Big|_{X_i=0}\sim {M_s^4\over m_P^2}\ ,}
which is
smaller by a factor of $\phi^2/m_P^2\sim10^{-10}$ than the contributions
(eq.~\xmass) we
have already considered.

\newsec{Discussion}
We have shown that it is possible to find an effective theory which dynamically
breaks
supersymmetry in a hidden sector, has a stable ground state with all field
expectation values well below the Planck mass, and which can naturally give
visible sector
gaugino masses of the same order as the squark and slepton masses. The model
does not
suffer from any of the cosmological problems associated with very light,
gravitationally
coupled scalars.   We see no experimental consequences  of this model other
than the
usual ones associated with supersymmetric extensions of the standard model---in
particular we do not see how to compute rather than estimate the superpartner
masses,
and this construction does not help explain the absence of flavor changing
neutral
currents\foot{If we assume the validity of string theory and  can find an
extension of
this construction which involves the superstring dilaton in the DSB dynamics,
we could
gain predictivity \ref\dilatonpredict{ V. S. Kaplunovsky, J.
Louis, Phys. Lett. B306 (1993) 269, hep-th/9303040; J. Louis,
Y. Nir, Nucl. Phys. B447 (1995) 18, hep-ph/9411429 }.}. Our model does
serve as an existence proof that the assumptions leading to the MSSM can be
realized in a
self-consistent effective field theory exhibiting dynamical supersymmetry
breaking.

\centerline{\bf Acknowledgements}
I would like to thank Michael Peskin for a conversation which inspired this
work,
Michael Dine for many useful comments, Joe Polchinski for a discussion about
supergravitational effects, and Yosef Nir and Ronen Plesser for sending me a
note about
their work in progress.
 This work   was supported in part by the DOE under grant
\#DE-FG06-91ER40614 and by the Alfred P. Sloan Foundation.

\appendix{A}{ Potential for Light Scalars} The classical scalar potential may
be
derived from the superpotential \nonrensuper.
The vaccuum state lies  along an  SU(7) D flat direction, with some field
strengths well above the scale $\Lambda_7$, where the  effective
theory has gauge group
 SU(5)  and  the following chiral superfields: a light
$10+\bar 5$,
$X_{1,2,3}$, and 3 SU(5) gauge singlet chiral superfields.
The effective theory    breaks
supersymmetry dynamically at the SU(5) scale.  If we consider
the effective theory at a short distance scale where it is weakly
coupled,
we can use
the classical potential derived from \nonrensuper\ to find the dilaton
and
$X_i$ scalar masses.  For simplicity we taking $\lambda_1=\sqrt2 =
\lambda_2=\sqrt2\lambda_3$, and expand about the D flat directions:
\eqn\susevendflat{\eqalign{
A=& \pmatrix{0&\phi/\sqrt6  & 0& 0& 0& 0& 0 \cr
  -\phi/\sqrt6 &0& 0& 0& 0& 0& 0 \cr
0&0& 0& 0& 0& 0& 0 \cr0&0& 0& 0& 0& 0& 0 \cr0&0& 0& 0& 0& 0& 0 \cr
0&0& 0& 0& 0& 0& 0 \cr0&0& 0& 0& 0& 0& 0 \cr}   \cr
 \bar F_1 =&{\phi\over\sqrt3}e^{i\alpha}
\pmatrix{ \cos\theta \cos\psi & -\sin\psi  & 0&0& 0& 0& 0}\cr
\bar F_2 = & {\phi\over\sqrt3}e^{i\beta}
\pmatrix{ \cos\theta \sin\psi& \cos\psi   & 0& 0&0& 0& 0}\cr
\bar F_3=& {\phi\over\sqrt3} \pmatrix{ \sin\theta&   0  & 0& 0& 0& 0& 0  },}}
($\theta,\psi,\alpha$ and $\beta$ are real). We find
the potential\eqn\effpot{\eqalign{
V_{\rm classical}=
&\abs{F_{X_1}}^2+\abs{F_{X_2}}^2+\abs{F_{X_3}}^2+\abs{F_{A_{12}}}^2
\cr &
+\abs{F_{\bar F_1^1}}^2+\abs{F_{\bar F_1^2}}^2 +\abs{F_{\bar F_2^1}}^2+
\abs{F_{\bar F_2^2}}^2+\abs{F_{\bar F_3^1}}^2+\abs{F_{\bar F_3^2}}^2
\cr &
{\rm (note\ that\ all\ other\ F\ terms\  are\ zero)}
\cr =&
 \abs{ \phi^6\left(1-{1\over 3}\cos(2\theta)\right)\over 9m^2_P}
 \cr & +
\abs{2\phi^4\over9 m^2_P}\Big\{ \abs{X_3}^2\left(1 + 2\cos^2\theta\right)
+ 2\abs{X_1}^2\left(1 +2\cos^2\psi\sin^2\theta\right)
\cr&
+ 2\abs{X_2}^2\left(1 + 2\sin^2\psi\sin^2\theta\right)
+4{\rm Re}\left(X_1X_2^* e^{i(\alpha-\beta)}\right)\sin(2\psi)\sin^2\theta
  \cr &
-2\sqrt2\sin(2\theta)
{\rm
Re}\left[X_3^*\left(X_1e^{i\alpha}\cos\psi+X_2e^{i\beta}\sin\psi\right)\right]
\Big\} \ . \cr}}
If classical physics were the whole story this potential would just drive
the theory to a supersymmetric state with
$\phi=0$. However    DSB   from the   SU(5) gauge theory gives  a
contribution
\eqn\sufivecont{
\Delta V\sim\Lambda_5^4\sim\abs{\phi}^{-16/13}\Lambda_7^{68/13}
}
 to the potential (note that eq.~\sufivecont\ is valid only for $\abs{\phi}\gg
\Lambda_7$), and the potential is minimized for $\phi\sim
\Lambda_7^{34/47}m_P^{13/47},\ \theta=0,\ X_{1,2,3}=0$, with all
noncompact
flat directions lifted.
Most scalars obtain mass of order
$\phi^2/m_P\sim M_s^2/\phi$ from the classical
superpotential, except those associated with 4 compact
flat directions of the potential, which are the result of the
spontaneous breakdown
of  exact global symmetries of the superpotential.
Either nonrenormalizable terms  in the K\"ahler potential, or still higher
dimension terms in the superpotential, or   supergravitational effects,
could explicitly break these global symmetries and lead to substantial
pseudo Goldstone boson
masses.
\listrefs
\bye
\end